\documentclass[aps,prl,twocolumn,showpacs,superscriptaddress,nobibnotes]{revtex4-1}

\usepackage{graphicx}
\usepackage[utf8]{inputenc}
\usepackage{siunitx}
\usepackage{amsmath}
\usepackage{amssymb}
\usepackage{enumerate}
\usepackage{color}
\usepackage{hyperref}
\usepackage[american]{babel}
\usepackage{etoolbox}
\usepackage{floatrow}
\usepackage{lineno,hyperref}
\usepackage{subfigure}
\usepackage[disable]{todonotes}
\usepackage{nicefrac}
\usepackage{booktabs}

\newcommand{\aegis}{AE$\bar{\hbox{g}}$IS}

\begin{document} 

\preprint{APS/123-QED}

\title{\LARGE\bf Absolute fraction of emitted Ps determined by GEANT4 supported analysis of gamma spectra}

\newcommand{\corresponding}[1]{\altaffiliation{Corresponding author, #1}}
\newcommand{\affcern}[0]{\affiliation{Physics Department, CERN, 1211~Geneva~23, Switzerland}}
\newcommand{\affinfnpd}[0]{\affiliation{INFN Padova, Via Marzolo~8, I-35131~Padova, Italy}}
\newcommand{\affinfnpdL}[0]{\affiliation{Department of Physics and Astronomy ''Galileo Galilei'', University of Padova, Via Marzolo~8, I-35131~Padova, Italy}}
\newcommand{\afffrmii}[0]{\affiliation{Heinz Maier-Leibnitz Zentrum (MLZ), Technische Universität München, Lichtenbergstrasse~1, 85748~Garching, Germany}}

\author{B.~Rien\"acker}
\corresponding{b.rienaecker@cern.ch}
\affcern
\afffrmii

\author{T.~Gigl}
\afffrmii

\author{G.~Nebbia}
\affinfnpd

\author{F.~Pino}
\affinfnpdL

\author{C.~Hugenschmidt}
\afffrmii

\date{\today}
      
\begin{abstract}
The fraction of positronium (Ps) emitted from a surface of a germanium single crystal at high temperature is usually assumed to approach unity at zero positron implantation energy.
In the experiment, however, the determination of the absolute Ps fraction is not straight forward since recording a reference spectrum with \SI{100}{\percent} Ps formation remains demanding.
We use GEANT4-simulated detector responses to $2\gamma$ and $3\gamma$ radiation sources mimicking positron and Ps annihilation inside the (coincidence) Doppler-broadening spectrometer at NEPOMUC, FRM II, in order to derive a reliable value for the Ps fraction reemitted from a Ge(100) target heated close to its melting point. 
Analysis of the measured spectra by fitting the simulated spectra shows an absolute value of $72\,\SI{+-4}{\percent}$ maximum Ps formation, contradicting the \SI{100}{\percent} assumption.

\end{abstract}

\maketitle{}

\section{Introduction}
Positronium (Ps) is a unique matter-antimatter hydrogen-like bound state of a positron and an electron. Positronium exists in two spin-configurations: the singlet state called para-Ps (p-Ps) with a total spin of $S=0$ and a vacuum lifetime of \SI{125}{\pico\second} annihilates predominantly into two $\gamma$ quanta with an energy of \SI{511}{\kilo\electronvolt} each \cite{Saito03}; the triplet state, called ortho-Ps (o-Ps), with $S=1$ and a ground state  lifetime in vacuum of \SI{142}{\nano\second} \cite{Koba08} annihilates predominantly into three $\gamma$ quanta of photon energy  $\leq$\SI{511}{\kilo\electronvolt} each, all three summing up to \SI{1022}{\kilo\electronvolt}.

Since its discovery in a nitrogen atmosphere by Deutsch in 1951 \cite{deutsch_ps:51}, Ps has flourished in a large variety of studies in fundamental research and applied sciences.
In solid-state and polymer physics, the free volume of amorphous matter can be characterized by the measurement of the o-Ps lifetime which is considerably reduced due to annihilation via the interaction of surrounding electrons (so-called pick-off process). 
Such investigations comprise, for example, the determination of the mean void size in polymers (see \cite{Jean} and references therein), as well as the characterization of systems with open porosity \cite{Gid06} and biopolymers \cite{Rou13, Hugen2014}. 
In fundamental research, Ps allows high-precision tests on bound-state quantum electrodynamics (QED) \cite{chu_mills:82,cassidy_paschen:11,cassidy_hyp:12,aegis_meta:18}. A limitation with respect to such experiments originates from higher-order effects that may introduce contributions from strong and weak interactions \cite{Karshenboim2004}. 
After the first observation of the Ps negative ion Ps$^-$ \cite{Mil81}, this leptonic three-particle system has served for high-precision QED tests as well \cite{Cee11b}.
Recently, first experiments demonstrated the route to generate a mono-energetic Ps beam via the production of Ps$^-$ and subsequent photo detachment of an electron \cite{Nag14}.
Experiments on antihydrogen currently underway at CERN, \aegis{} \cite{KELLERBAUER2008351} and GBAR \cite{Perez12}, make use of Ps and its ability to enhance the antihydrogen production via charge exchange. Indeed, \aegis{} recently succeeded in producing the first pulsed antihydrogen source using Rydberg excited Ps and cold antiprotons \cite{aegis19}. 

Research with Ps clearly benefits from intense cold Ps sources (around one-hundred \SI{}{\milli\electronvolt}) based on efficient conversion of positrons into positronium. Positrons are usually provided by radioactive sources like $^{22}$Na or via pair production as exploited at the NEutron induced POsitron source MUniCh (NEPOMUC). 
Formation of Ps can occur at surfaces where its emission is mainly thermally activated involving positrons trapped in surface states. 
At surfaces of metals and semiconductors, high values of Ps formation at high temperature and $\sim$\SI{0}{\electronvolt} positron implantation energy have been measured \cite{Mills:1978, Lynn:1980, Nieminen:1980, Soininen:1991, Cass:2011, Nagashima2018}. 
In bulk, however, Ps-formation is suppressed due to the effective shielding of conduction electrons.
Insulators, on the other hand, and in particular porous ones, allow positron-electron binding inside the bulk and subsequent emission into vacuum with efficiencies in the range 45-\SI{84}{\percent} of the amount of implanted positrons \cite{Eldrup1985,Nagashima95,mariazzi_Ps:10,Petegem04}.
Therefore, efficient Ps converters are often nanoporous targets where positrons are implanted, thermalize and diffuse back to the surface. Either during the diffusion inside the bulk or directly at the surface, positrons can bind to an electron and be emitted as Ps.

In general, there are two methods by which the amount of free Ps is estimated. 
The first technique  exploits the characteristic long lifetime of o-Ps by recording time-resolved spectra after an initial pulsed production with a narrow time spread. 
The delayed annihilation signal of ground-state o-Ps in flight thus enables an estimation of the amount of Ps produced \cite{CASSIDY20071338,Cass11:SSPALS}. 
This method, however, can only serve as a rough estimate for the emitted Ps fraction due to several systematic uncertainties such as limited flight space, non-linear detector responses and saturation effects at short times, where the signal is dominated by the initial peak of positron annihilation.

The second approach is the detection of annihilation radiation with high energy resolution of Ps either self-annihilating in flight or annihilating upon interaction with an obstacle.
This technique ultimately aims to separate the amount of  $2\gamma$ and $3\gamma$ annihilation events serving as fingerprints of positron/p-Ps and o-Ps  annihilations, respectively. 
In principle, the relative change of Ps emission from different samples or by variation of external parameters like positron implantation energy or temperature can easily be estimated from the recorded energy spectra, whereas the determination of the absolute Ps fraction is not straight forward.
A commonly used approach is the comparison of the measured $\gamma$ spectrum with a reference spectrum recorded for 100\% Ps formation.
By referring to an experiment performed by Mills \cite{Mills:1978} it is often assumed that positron implantation with various energies and extrapolating to zero energy yields 100\% Ps formation in germanium at high temperature, while other loss channels such as positron annihilation from a surface state can be neglected.
Such a spectrum showing 100\% Ps formation, however, is demanding to be obtained experimentally and cannot easily be verified. On the contrary, a theoretical calculation \cite{Shrivastava:1990} even finds a saturating upper limit of \SI{90}{\percent} for the Ps emission from Ge(110) at high temperatures.

In the present paper we discuss the various drawbacks of the commonly applied analysis method. 
Annihilation spectra that are dependent on positron implantation energy have been recorded for a Ge(100) target close to its melting point.
By including the relevant parts of the spectrometer, we simulated the $\gamma$ spectra for $2\gamma$ and $3\gamma$ annihilation in order to reliably estimate the absolute fraction of thermal and non-thermal Ps emitted from a Ge(100) surface.

\section{Ps emission from surfaces}

\subsection{Positrons and Ps at surfaces}

    The one-dimensional distribution of positrons thermalized in the bulk of a solid as a function of the incident positron energy can be described by the Makhovian implantation profile. The fraction of thermalized positrons at depth $z$ with respect to the surface for a positron energy $E$ is then given by material-dependent parameters  $n,m,B$ and $z_0=BE^n$ [see Eq.\,\ref{eq:diffusion_model}] which can be found, for example, by means of Monte Carlo simulations \cite{Dryzek:2008}. After thermalization, the positrons obey the laws of diffusion. Multiplying the term for Makhovian positron implantation with the exponential term for subsequent positron diffusion and integrating the result from zero to infinity returns the amount of positrons diffusing back to the surface $J(E)$ \cite{Schultz:1988}:
    \begin{equation} \label{eq:diffusion_model}
		J(E) = J_0\int_0^\infty \dfrac{m}{{z_0}^m}z^{m-1} e^{-\big(\tfrac{z}{z_0}\big)^m}e^{-\tfrac{z}{L_{+}}}dz .
	\end{equation}
    The surface coefficient $J_0$ denotes the maximum fraction of thermalized positrons that is able to diffuse back to the surface and finally escape from the target. This escape can either occur by Ps production without trapping, by reemission due to a negative positron work function or by trapping into a surface state. Internal reflection at the surface and subsequent reimplantation into the bulk, on the other hand, is a loss channel for positrons causing $J_0$ to be smaller than one. In order to obtain highest Ps yield, the coefficient $J_0$ has to approach unity, which is only given at high sample temperatures \cite{Nieminen:1980}. 
    
    The relation between the amount of emitted Ps and thermalized positrons is then given by:
    \begin{equation} \label{eq:fPswoBG}
	    f_{Ps}(E) = f_0J(E),
    \end{equation}
    where $f_0$ is the fraction of Ps that can be emitted from the surface with respect to all thermalized positrons arriving at the surface. By accounting for the different positron escape rates $\nu_{i}$ at the surface that involve thermal positrons, $f_0$ can therefore be expressed by the ratio 
    \begin{equation} \label{eq:f0_nu}
        f_0 = \dfrac{\nu_{s,Ps} + \nu_{0,Ps}}{\nu_{s,Ps} + \nu_{0,Ps} + \nu_{0,e^+} + \nu_{s,2\gamma}} 
    \end{equation}
    Here, $\nu_{s,Ps}$ denotes the rate of positrons getting trapped in a surface state and being thermally desorbed as Ps, and $\nu_{0,Ps}$ is the rate of positrons that bind to a surface-electron without trapping. The other two escape channels, $\nu_{s,2\gamma}$ and $\nu_{0,e^+}$, are the rate of positrons getting trapped in a surface state and annihilating with electrons into $2\gamma$ quanta, and the re-emission of positrons into vacuum, respectively \cite{Nieminen:1980}. The latter case is only possible, if the material has a negative positron work function.
    
    From the two Ps emitting channels, only the thermal desorption leads to thermal Ps. The other, $\nu_{0,Ps}$, depends on the value of a negative Ps work function, which, if occurring, generally lies around few \SI{}{\electronvolt} \cite{Nagashima98}. However, not fully thermalized (epithermal) positrons can also contribute to the non-thermal Ps emission, but these are not correctly described by the diffusion model and therefore have to be treated independently. As a remark, epithermal positrons may also escape directly from the surface if their energy is exceeding the (positive) positron work function.  

    In an experiment aiming to obtain \SI{100}{\percent} Ps formation, one has to enable only those positron escape routes that produce Ps, while the others ($\nu_{0,e^+}$, $\nu_{s,2\gamma}$ and epithermal positron emission) have to become negligible. 

\subsection{Analysis of energy-resolved positron annihilation spectra}

\begin{figure}[thpb]
        \centering
        \includegraphics[width=1 \linewidth]{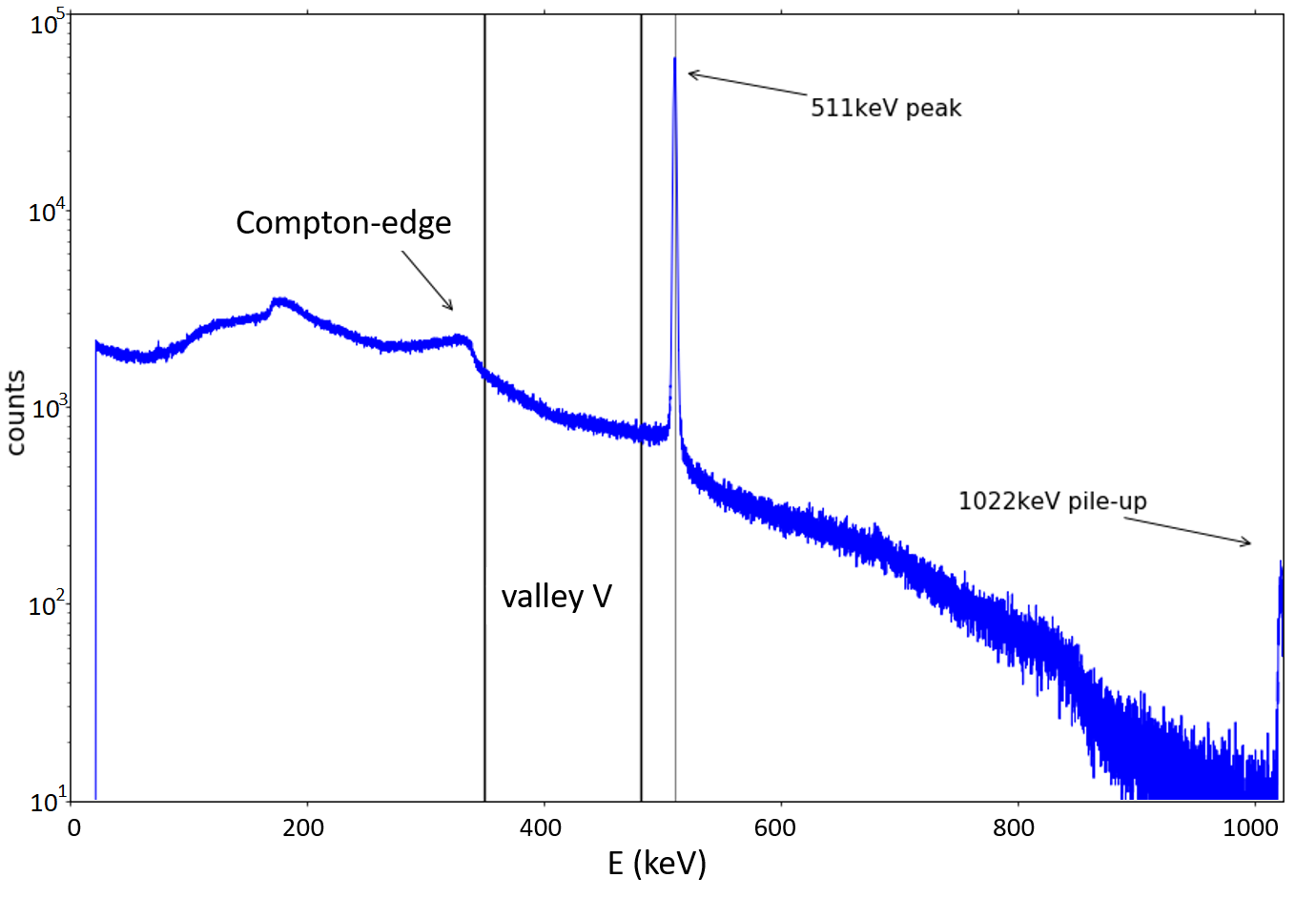}
        \caption{Exemplary gamma spectrum resulting from positrons implanted at \SI{15}{\kilo\electronvolt} detected by a high-purity germanium detector. The important features of this spectrum are the main positron annihilation photo peak at \SI{511}{\kilo\electronvolt}, the pile-up peak at the right edge due to simultaneous detection of two such gamma quanta (\SI{1022}{\kilo\electronvolt}), the Compton edge at \SI{340}{\kilo\electronvolt}, and the valley-region V, which is the most sensitive to $3\gamma$ annihilation of o-Ps.}
        \label{fig:whole_CDBspec}
    \end{figure}

    The analysis of energy-resolved $\gamma$-spectra as for example obtained with high-purity germanium (HPGe) detectors will be briefly reviewed using a spectrum recorded for a Ge-target with the (coincidence) Doppler-broadening spectrometer (CDBS) in Munich \cite{Gigl17} at \SI{15}{\kilo\electronvolt} positron implantation energy. In Fig. \ref{fig:whole_CDBspec}, the abscissa denotes the photon energy recorded by a multichannel analyzer, whose channels have been converted into energies via calibration with an $^{152}$Eu source. For such a spectrum, a so-called valley-region can be defined as marked in the figure,  which is the most sensitive region to o-Ps annihilation due to the low background. Other contributions in this region are the environmental background, pile-up events (coincident detection of two photons with lower energy), and annihilation gammas experiencing small-angle Compton scattering inside the sample or other materials in the field of view of the detector. 
 
    By switching off the positron source, the environmental background can simply be measured and subtracted from any spectrum of interest. 
    The pile-up and small-angle Compton scattering events can roughly be approximated, for example, by a sigmoidal function, and also be subtracted - or sometimes they are simply considered negligible.
    
    In order to estimate the produced amount of Ps, one starts by defining the so-called valley-to-peak ratio $R=\frac{V}{P}$, where $V$ is the integral of the valley-region and $P$ is the integral of a small window around the photo-peak at \SI{511}{\kilo\electronvolt} after background subtraction, respectively. Sometimes $R$ is also defined as $R=\frac{T-P}{T}$, with $T$ being the  total integrated spectrum. In both cases, the term $f_{Ps}$ (also called the $F$-parameter) can be derived from a spectrum using the value $R$ \cite{Marder:1956,Mills:1978}:
    \begin{equation} \label{eq:F}
		f_{Ps} = \bigg(1 + \dfrac{P_1}{P_0}\cdot\dfrac{[R_1-R]}{[R-R_0]}\bigg)^{-1}
	\end{equation}
	The subscripts 0 and 1 indicate that the corresponding values have to be obtained from two reference measurements with \SI{0}{\percent} and \SI{100}{\percent} Ps formation, respectively.
	One has to make sure that the values of $P_{0,1}$ are obtained with the same amount of implanted positrons, since there is no automatic normalization included in this ratio (unlike the R-values). 
	However, this approach has to deal with a number of other possible error sources which are summarized in the following.
	
	A first uncertainty originates from the experimental setup, which influences the efficiency and the amount of detected gamma quanta. Naturally, only a fraction $<1$ of total Ps emission can be detected, for example due to gamma-absorption in the surrounding material. For this reason, one cannot simply verify the absolute amount of produced Ps. 

	The velocity distribution of Ps leaving the target leads to a slow fraction annihilating in flight close to the origin and a faster one rather far from it \cite{Jorch:1984,Dupas:1985}. Hence, in an experiment the geometric acceptance of the detector will weight the occurring radiation differently depending on the location of an annihilation event. Limiting the flight path of o-Ps, for example with an obstacle, will lead to pick-off annihilation which produces $2 \gamma$'s and the event is wrongly sorted into the peak region. The same holds for fast o-Ps actually hitting the chamber walls.
	
	A generally trusted method for obtaining a \SI{100}{\percent} reference is to implant positrons into a pure germanium single-crystal heated close to its melting point (\SI{1210}{\kelvin}) and extrapolate the recorded data to zero kinetic energy \cite{Mills:1978}.
	Although an upper limit for Ps conversion efficiency of \SI{97}{\percent} has been achieved with a Ge target using positron-annihilation induced Auger-electron spectroscopy (PAES) as an independent verification \cite{Soininen:1991}, one usually just assumes that the obtained Ps amount following from the Ge measurement is close to \SI{100}{\percent}, without further validation. However, surface contamination and the difficulty to prepare a positron beam with very low kinetic energy are important factors that directly challenge this assumption. Additional factors are backscattered or reemitted positrons and unwanted annihilations in the bulk due to impurities and other defects.

    But also the assumption of \SI{0}{\percent} Ps formation is not without risk - even for high implantation energies - due to the energy and material-dependent back-scattering probability of positrons, which ranges between \SI{12}{\percent} to \SI{25}{\percent} for germanium \cite{Makinen92}. This renders a considerable fraction of positrons to end up in an unpredictable state. In addition, large vacancy clusters and pores can lead to unwanted Ps formation in the bulk of the sample.

\section{Experiment}

\begin{figure}[thpb]
        \centering
        \includegraphics[width=1 \linewidth]{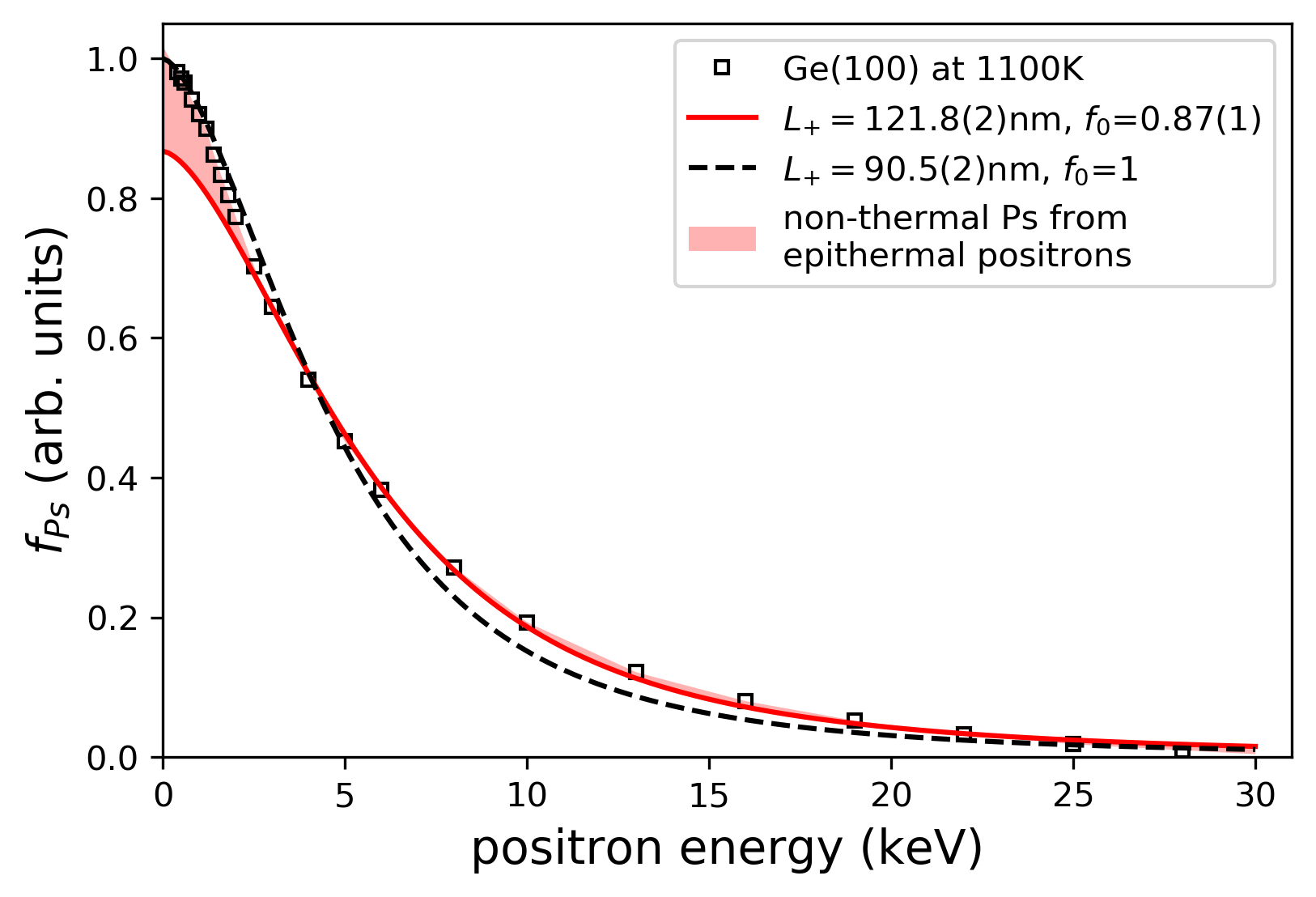}
        \caption{Ps yield from pure Ge(100) heated to \SI{1100}{\kelvin} as a function of positron implantation energy using the traditional F-parameter formalism as specified in the text. Measured values (symbols) were used for fitting models describing the amount of implanted positrons diffusing back to the surface and forming Ps. The term $L_{+}$ is the positron diffusion length and $f_0$ the maximum fraction of Ps formation following from thermal positron diffusion towards the surface. Note that $f_0$ can contain components of thermal and non-thermal Ps. Statistical error bars on the data are smaller than the symbols.}
        \label{fig:OldFGe1100K}
    \end{figure}

	We fixed a piece of an undoped Ge(100) wafer onto a molybdenum plate, which was attached to a controllable heater and a temperature sensor, all in an ultra high vacuum of \SI{5e-8}{\milli\bar}.
	The sample was heated to \SI{870}{\kelvin} for about \SI{2}{h} in order to clean the Ge surface of remaining contaminants and to anneal it. Then, the temperature was increased to \SI{1100}{\kelvin}, similar to the procedure described in \cite{Soininen:1992}. 
	For our experiments, we used the 20\,eV positron beam with an intensity of \SI{5e7} remoderated positrons per second from the positron source NEPOMUC. Inside the coincidence Doppler-broadening spectrometer, the positron beam was accelerated and  implanted into the sample with adjustable energies ranging from \SIrange{0.4}{28}{\kilo\electronvolt} \cite{Hugen15,Gigl17}.
	An environmental background in the absence of positrons was acquired for \SI{600}{\second}, the Ge-spectra were acquired for \SI{300}{\second}. The latter resulted in \SI{7e6}{} detected counts on average when integrating over the whole spectrum.	
	The Ge spectra were evaluated for all energies using Eq. \ref{eq:F}, still assuming that at \SI{0.4}{\kilo\electronvolt} close to \SI{100}{\percent} of free Ps is produced and \SI{0}{\percent} at \SI{28}{\kilo\electronvolt}. 
	Although it is obvious that highest yield of free Ps can be expected at lowest positron implantation energy, the assumption of \SI{100}{\percent} free Ps is - as reasoned above - not justified \textit{a priori}. Therefore, the experimental data in Fig. \ref{fig:OldFGe1100K} are shown in arbitrary units, i.e. the relative change of Ps formation with respect to the value at \SI{0.4}{\kilo\electronvolt} is correctly given. The absolute surface value of $f_{Ps}$, however, has to be verified by a simulation as presented in Section \emph{Simulation}. 
	
    Using the diffusion model presented in Eqs. \ref{eq:diffusion_model} and \ref{eq:fPswoBG} for Ge with the parameters $n=1.6$, $m=1.68$, $B=\SI{7.48}{\nano\meter\,\kilo\electronvolt^{-n}}$ from \cite{Dryzek:2008} and using $f_0=J_0=1$ (assuming that only thermal positrons contribute to $\SI{100}{\percent}$ Ps production), one can numerically fit this model to the experimental data via the least-squares method with diffusion length $L_{+}$ as the free parameter. The result is shown in Fig. \ref{fig:OldFGe1100K} as the dashed black line.
    The term $J(E)$ can additionally be evaluated at $E=\SI{0.4}{\kilo\electronvolt}$ with the obtained diffusion length of \SI{90}{\nano\meter}, which results in \SI{99}{\percent} of thermal positrons returning to the surface. This confirms the initial assumption of $J_0$ = 1, but it does not tell anything about $f_0$, which denotes the final escape route of the positrons. 
    Furthermore, this fitted curve does not resemble well the measured data at all energies. Indeed, the extracted diffusion parameter $E_0$ (connecting the diffusion length with the Makhovian material constants by the relation $L_{+} = B{E_0}^n$) amounts to \SI{4.8+-0.1}{\kilo\electronvolt}, which does not correspond to earlier results, obtained for example in Ref. \cite{Soininen:1992} where $E_0$ for Ge at \SI{1100}{\kelvin} was around \SI{6}{\kilo\electronvolt}. Changing the fit procedure in a way that $f_0$ is a fit parameter, too, while keeping $J_0=1$ and using only energies $>\SI{2}{\kilo\electronvolt}$ in order to avoid contributions from not fully thermalized positrons that are not obeying the diffusion law \cite{Huomo89}, one obtains an improved fit (solid line) with $f_0=\SI{0.87+-0.01}{}$ and $L_{+}=\SI{121.8+-0.2}{\nano\meter}$ (i.e. $E_0 = \SI{5.8+-0.1}{\kilo\electronvolt}$). This indicates that the contribution of nonthermal Ps to the total amount of emitted Ps is at least \SI{13}{\percent}.
  
    In the next paragraph we introduce a GEANT4-supported model allowing us to estimate the Ps-emission. GEANT4 is a well-established toolkit for simulating the passage of particles through matter (see Ref. \cite{GEANT4} for more details).
    We simulated detector responses to $3\gamma$ and $2\gamma$ annihilation sources as if they were located inside our experimental setup at the target position, including the relevant photon interactions (Rayleigh scattering, Compton scattering, photoelectric effect and photon absorption within surrounding materials).
    A superposition of the simulated detector responses for $3\gamma$ and $2\gamma$ annihilation events is then fitted to measured spectra in order to determine the absolute Ps yield. 

\section{Simulation} \label{secsimu}

A pure Ps annihilation spectrum consists of $N_{3\gamma}$ photons from $3\gamma$ annihilations (continuous spectrum up to \SI{511}{\kilo\electronvolt} as derived by Ore and Powell \cite{OrePowell49}), and of $N_{2\gamma}$ photons from $2\gamma$ annihilations (discrete energy of $\SI{511}{\kilo\electronvolt}$ per photon). The formation ratio of ortho-Ps to para-Ps is 3:1 according to the spin multiplicity of the triplet state. Hence, the respective number of photons of a pure Ps spectrum has to obey the ratio $N_{3\gamma}$:$N_{2\gamma}$ = 9:2. Consequently, knowing the amount of $3\gamma$-events in a measurement is equivalent to the knowledge of the number of ortho-Ps atoms. This in turn reveals the total fraction of produced Ps, i.e. p-Ps and o-Ps, with respect to the total number of implanted positrons:
\begin{equation} \label{eq:fPs}
	f_{Ps} = \dfrac{\text{amount of Ps}}{\text{total number of $e^+$}} = \dfrac{\tfrac{4}{3}N_{3\gamma}}{N_{3\gamma} + \tfrac{3}{2}N_{2\gamma}}
\end{equation}

As the detector exhibits a unique energy-dependent response to gamma radiation, it is not trivial to count the number of respective quanta from a measured spectrum like in Fig. \ref{fig:whole_CDBspec}. One approximate solution was presented in Ref. \cite{Hugenschmidt08}, where the integrated valley region $V$ as defined above (most sensitive to $3\gamma$ photons) was scaled according to the full theoretical $3\gamma$-spectrum derived by Ore and Powell \cite{OrePowell49} and was put into ratio to the total integrated signal, yielding an estimation for $f_{Ps}$.

\begin{figure}[thpb]
    \centering
    \includegraphics[width=0.85 \linewidth]{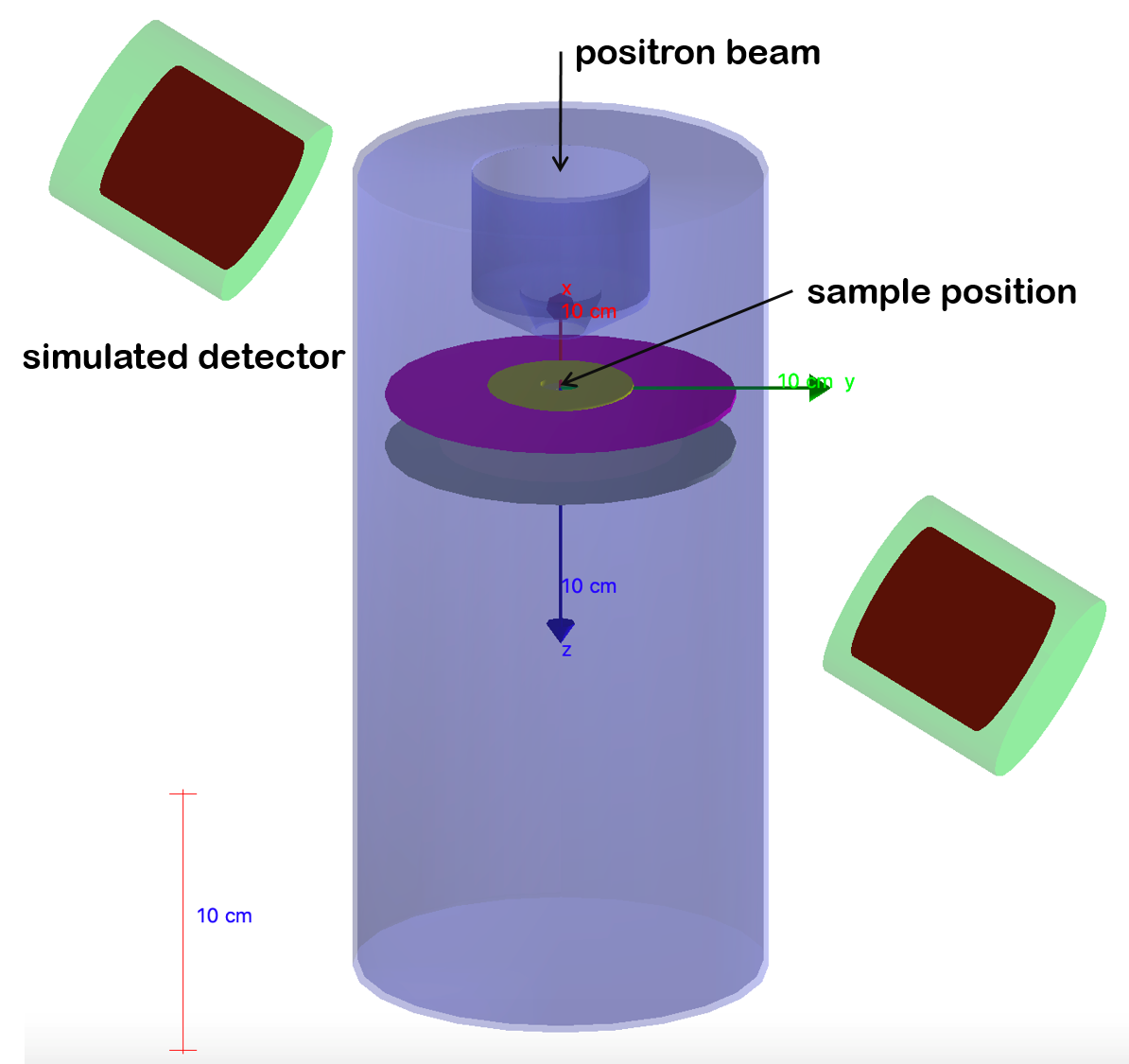}
    \caption{GEANT4-applicable sketch of the (coincidence) Doppler-broadening spectrometer at NEPOMUC as it was set up during the measurements on Ge(100) targets. In the center of the upper molybdenum disk, where the sample was located during the experiment, a point source of annihilation radiation is assumed, mimicking a pure $2\gamma$-spectrum and a continuous $3\gamma$ spectrum, respectively. Spectra as usually recorded by the surrounding upper detector were therefore simulated.}
    \label{fig:CDBsketch}
\end{figure}

In order to take into account the unique detector response to different energies as well, we introduced the geometry of the experimental setup including the HPGe detector into a GEANT4 simulation (Fig. \ref{fig:CDBsketch}). By this, major physical effects such as the Compton-effect or gamma absorption by solids are appropriately taken into account. In the Ps formation region, the CDB spectrometer consists of a cylindrical sample chamber made of stainless steel with \SI{2}{\milli\meter} thick walls and \SI{150}{\milli\meter} inner diameter, two \SI{2}{\milli\meter} thick aluminum rings (working as electrostatic lenses set to high potential) with an outer radius of \SI{65}{\milli\meter}. A molybdenum disk as part of the target heater was placed in the center of the upper aluminum ring. At the top, roughly \SI{60}{\milli\meter} away from the upper aluminum disk, is the last steel electrode of the positron beam guiding system, allowing to steer the positrons directly onto the target placed below. 
Outside the sample chamber, four HPGe detectors are located facing towards the center of the molybdenum disk, where the Ge(100) target was located. Finally, only the shown upper one was actually used to simulate the detector response. The detector itself was modeled as a cylinder of solid germanium with radius \SI{28}{\milli\meter} and a length of \SI{54}{\milli\meter}, encapsulated inside an aluminum shell with \SI{1}{\milli\meter} thin walls. The detector front face was \SI{140}{\milli\meter} away from the center between the two disks. Exact details of the geometrical and functional setup can be found in Ref. \cite{Gigl17}.

In the simulation, we assumed a point source in the center of the heater from where annihilation radiation is emitted isotropically. First, we used the raw $3\gamma$ spectrum corresponding to pure o-Ps annihilation. Then, we introduced a $2\gamma$ source which generated two photons with \SI{511}{\kilo\electronvolt} energy emitted in opposing directions. The energy-dependent energy resolution of the HPGe detectors have been found experimentally via the aforementioned $^{152}$Eu calibration measurements. This information is required for the convolution of the simulated spectra with the finite energy resolution of the detectors.

\begin{figure}[thpb]
    \centering
    \includegraphics[width=1 \linewidth]{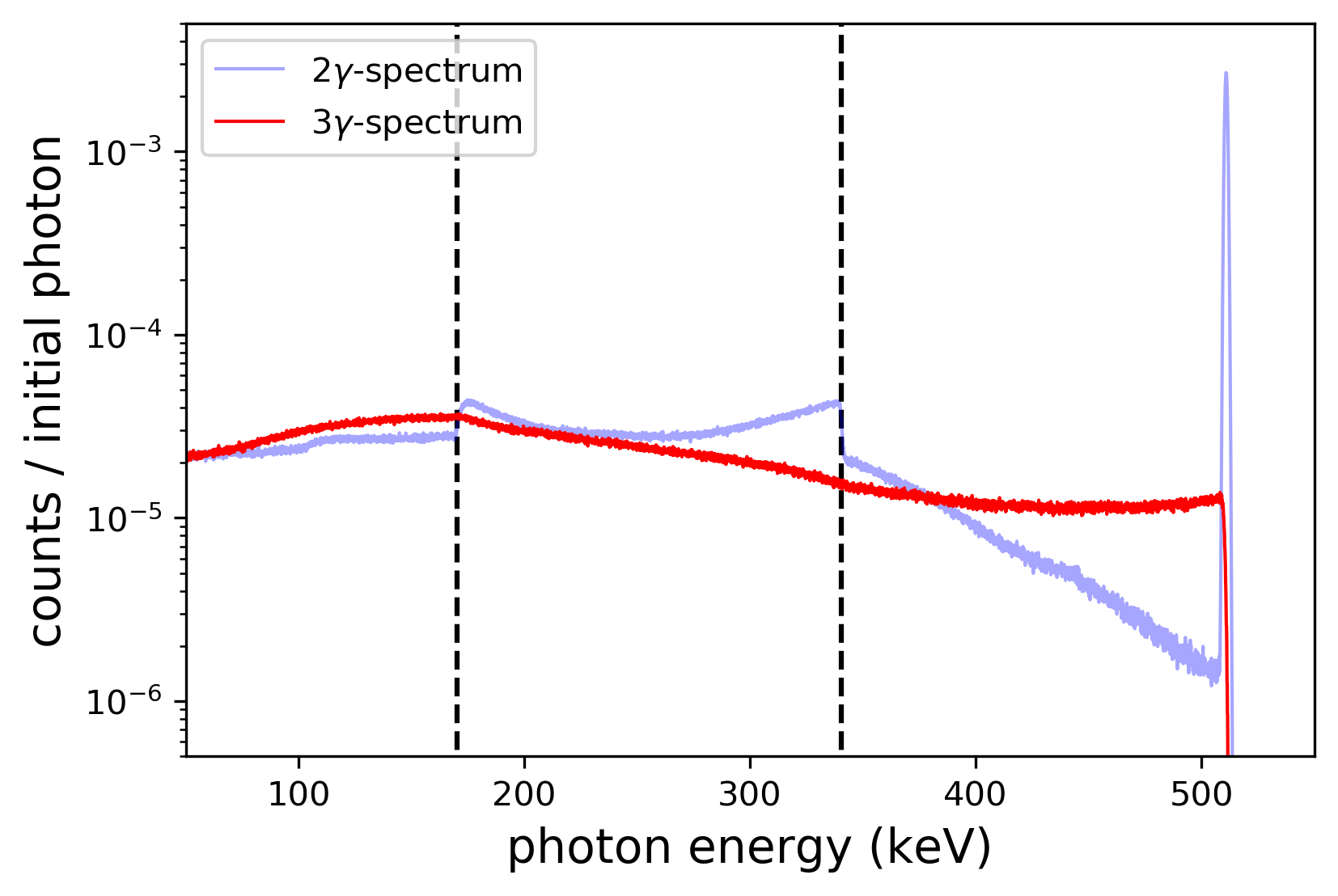}
    \caption{Simulated Ps annihilation spectra for the upper HPGe detector in the CDB spectrometer. Light blue (grey): pure $2\gamma$ spectrum as from para-Ps or direct positron annihilations. Red (black): $3\gamma$ spectrum as from ortho-Ps annihilations. The dashed lines mark the energies \SI{170}{\kilo\electronvolt} and \SI{340}{\kilo\electronvolt}, hence $\tfrac{1}{3}$  and $\tfrac{2}{3}$ from the photo peak energy.}
    \label{fig:sim3g2g}
\end{figure}

The two spectra following from the simulation with \SI{1e9}{} initial photons are very distinct as can be seen in Fig. \ref{fig:sim3g2g}. Both spectra were normalized to the same number or photons. For the $2\gamma$ spectrum $C_{2\gamma}$, we find the main peak at \SI{511}{\kilo\electronvolt}, the Compton-edge at \SI{340}{\kilo\electronvolt} and an edge at \SI{170}{\kilo\electronvolt} originating from photons Compton-scattered in the surrounding material (e.g. the steel chamber walls, where a full backscattering event would allow to reach the detector in the opposite direction, while the photon has lost $\tfrac{2}{3}$ of its original energy). The $3\gamma$ spectrum $C_{3\gamma}$ is rather flat, lacking as expected a clear peak-structure. 
Finally, one can find the values for $N_{3\gamma}$ and $N_{2\gamma}$ by fitting a linear combination of both spectra to any measured data with our detector-system:
\begin{equation} \label{eq:fitsim}
	C_{meas}(E) = N_{3\gamma}C_{3\gamma} + N_{2\gamma}C_{2\gamma} + C_{BG},
\end{equation}
where $C_{BG}$ is a positron-induced background that might be present in the measured spectrum, as detailed in the following section.

\section{RESULTS AND DISCUSSION}

In order to evaluate the total fraction of produced Ps, $f_{Ps}$, from the same experimental data shown in Fig. \ref{fig:OldFGe1100K} we applied the procedure described above utilizing the simulated spectra. 
First, each measured spectrum is normalized to the same measurement time resulting in a constant integrated positron signal, since positrons were implanted at a constant positron rate. 
Second, the environmental background obtained with the positron beam switched off is subtracted from all spectra.
The positron induced background $C_{BG}$ (see Eq.\,\ref{eq:fitsim}) is found by taking the difference between the simulated $2\gamma$ spectrum and a measured spectrum where essentially no Ps-formation occurs, i.e. with only $2\gamma$ annihilations and subsequent scattering processes. 
For this, a spectrum on Ge(100) at high implantation energy (\SI{28}{\kilo\electronvolt}) and low target temperature (\SI{300}{\kelvin}) was recorded. As before, the expected fraction of back-diffusing positrons to the surface at this high energy can be estimated via Eq. \ref{eq:diffusion_model} using the positron diffusion length found in the experimental section. This fraction shows to be negligible, allowing to use the corresponding spectrum as a \SI{0}{\percent} Ps reference.

\begin{figure}[thpb]
    \centering
    \includegraphics[width=1 \linewidth]{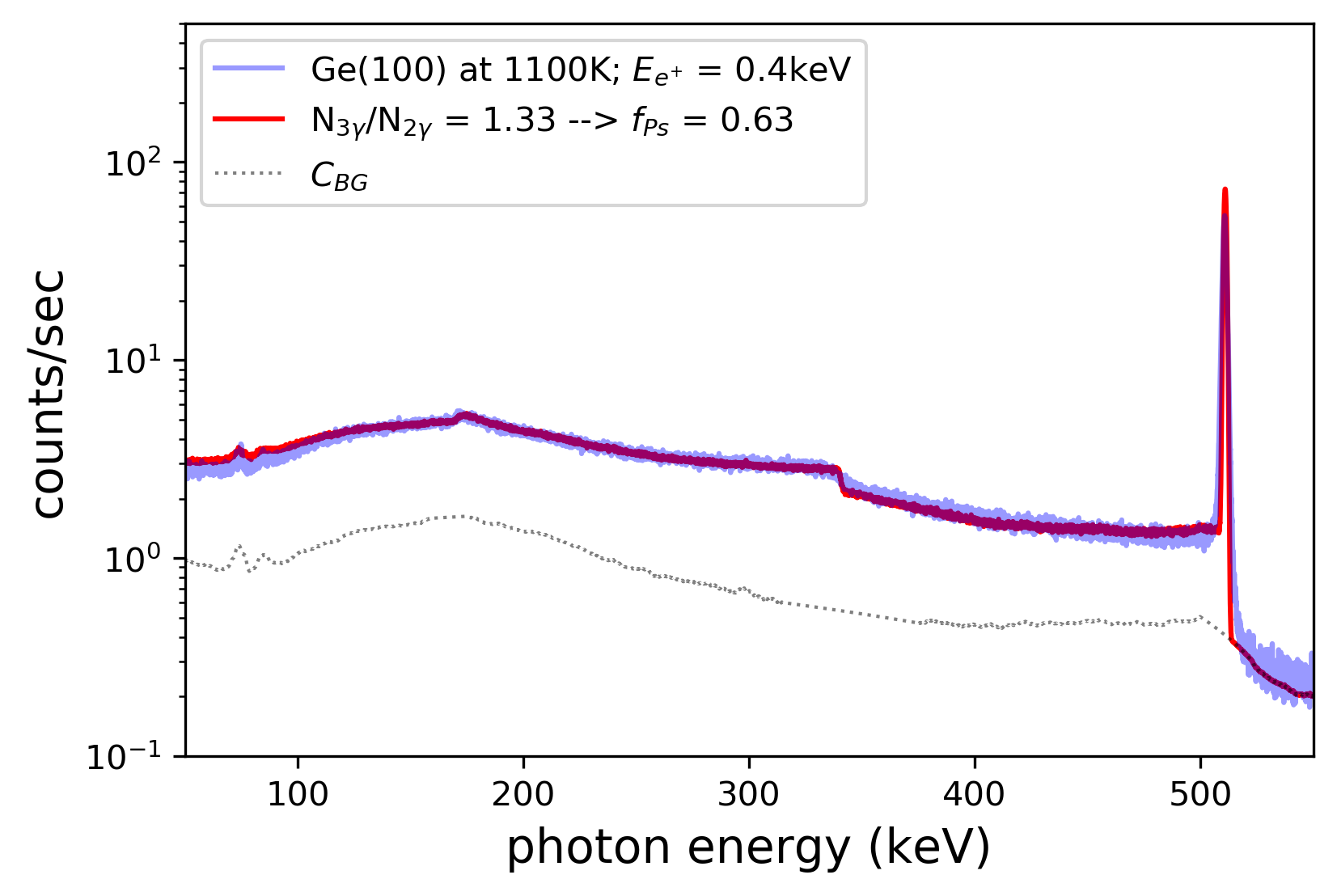}
    \includegraphics[width=1 \linewidth]{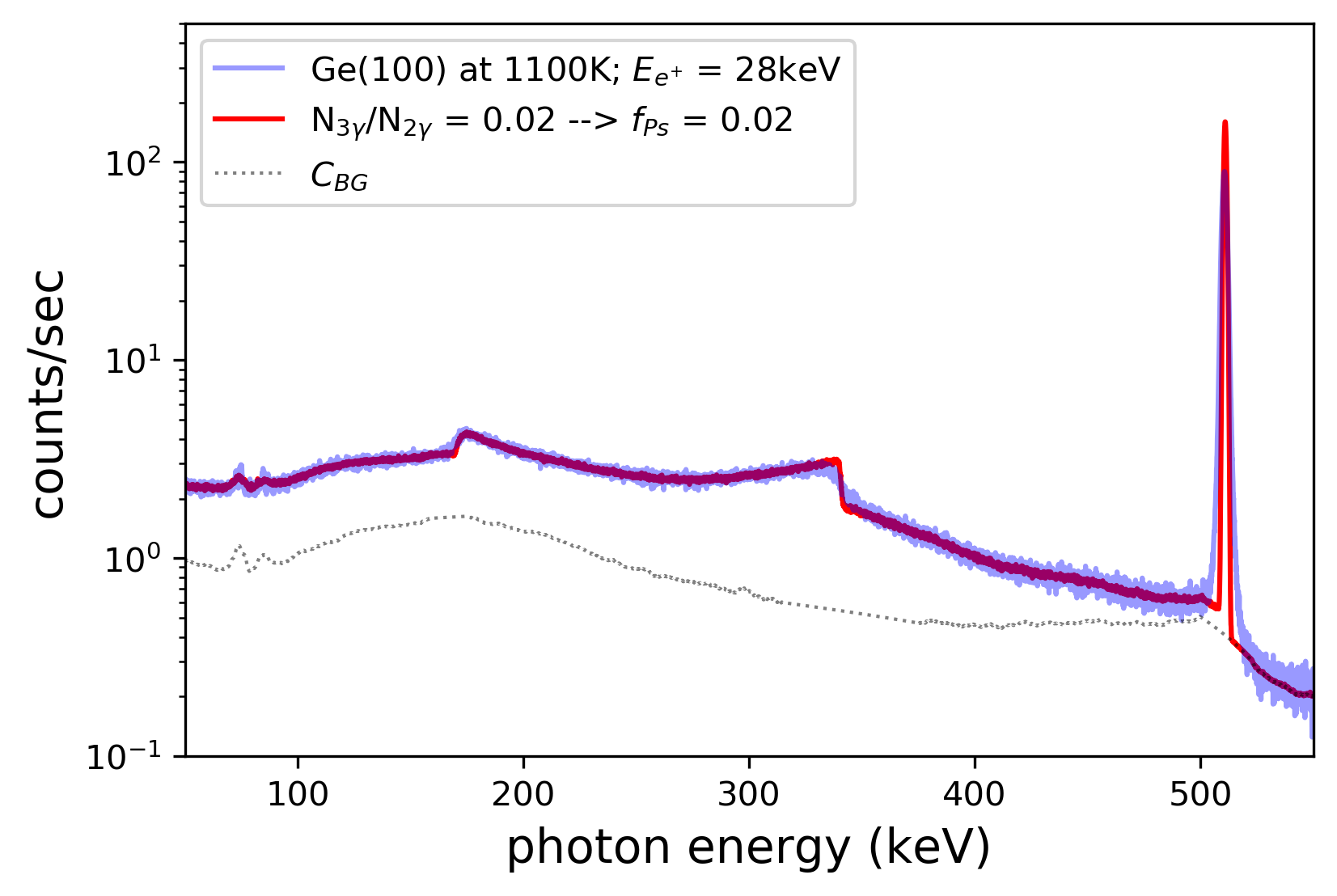}
    \caption{Measured spectra (light blue/gray) fitted with a superposition of simulated $2\gamma$- and $3\gamma$-spectra (red/black) for a positron implantation energy of $\SI{0.4}{\kilo\electronvolt}$ (top)
    and $\SI{28}{\kilo\electronvolt}$ (bottom). The positron-induced background $C_{BG}$ (dotted line) is the baseline on top of which the superimposed simulated spectra are derived. The amount of emitted Ps $f_{Ps}$ is derived from the ratio $N_{3\gamma}/N_{2\gamma}$, yielding 0.63 and 0.02, respectively.}
    \label{fig:Gefitted}
\end{figure}

We scaled the simulated $2\gamma$ spectrum $C_{2\gamma}$ by taking the integral within an energy interval of $\pm{2}\sigma$ of the photo peak to be equal to the one of the measured spectrum. 
Consequently, the difference between measured and scaled simulated spectrum yields the background spectrum $C_{BG}$ as shown in Fig. \ref{fig:Gefitted}. Artifacts due to the subtraction were smoothed via moving window averaging.

The background spectrum $C_{BG}$ is subtracted from all other measured spectra in order to obtain the "true" annihilation spectrum which can be described by a superposition of a pure $2\gamma$ spectrum and a pure $3\gamma$ spectrum. 
First, the weighting factor of the $2\gamma$ spectrum has been extracted by scaling the simulated $C_{2\gamma}$ in such a way that its peak integral within $\pm{2}\sigma$ around the peak center matches that of the measured one. 
After subtracting this scaled $C_{2\gamma}$ spectrum, the simulated $C_{3\gamma}$ spectrum was scaled up such that its integral between \SI{100}{\kilo\electronvolt} and \SI{500}{\kilo\electronvolt} photon energy matched the one calculated for the remainder. The two determined scaling factors correspond to $N_{2\gamma}$ and $N_{3\gamma}$ in Eq. \ref{eq:fitsim}, respectively, from which $f_{Ps}$ can be calculated.
An example for the measurement with positrons implanted in Ge(100) at \SI{1100}{\kelvin} with \SI{0.4}{\kilo\electronvolt} and \SI{28}{\kilo\electronvolt} is shown in Fig. \ref{fig:Gefitted}. The ratio of $N_{3\gamma}/N_{2\gamma}$ is found to be $1.33$ and $0.02$, respectively. This can be translated via Eq. \ref{eq:fPs} to the total fraction of Ps production $f_{Ps} = 0.63$ and $f_{Ps} = 0.02$.

It is noteworthy that without subtracting the background $C_{BG}$, we find up to \SI{77}{\percent} Ps formation at our lowest implantation energy, while $f_{Ps}$ asymptotically approaches $\SI{25}{\percent}$ for increasing energies. This behavior cannot be explained through Ps formation, since this escape route is strongly suppressed in the high-energy regime. 
However, the detected counts at energies higher than \SI{511}{\kilo\electronvolt} indicate significant contributions due to pile-ups. Mainly dependent on the detection system, such pile-ups are falsely counted as $3\gamma$ annihilation events when occurring on the left-hand side of the photo peak. In addition, the positron induced-background comprises effects which the simulation does not account for such as backscattered positrons. 
This is actually the reason for including the term $R_0$ in the traditional $F$-parameter calculation as denoted in Eq. \ref{eq:F}, which consists of all the background events that can not reasonably ascribed to Ps formation. This is of particular importance since one would falsely obtain more than \SI{50}{\percent} Ps-formation by setting $R_0=0$ using for example the results reported in Ref. \cite{Huomo89}.
In conclusion, it is reasonable to use the described background subtraction in order to find a trustworthy lower limit for the amount of produced Ps.

\begin{figure}[thpb]
    \centering
    \includegraphics[width=1 \linewidth]{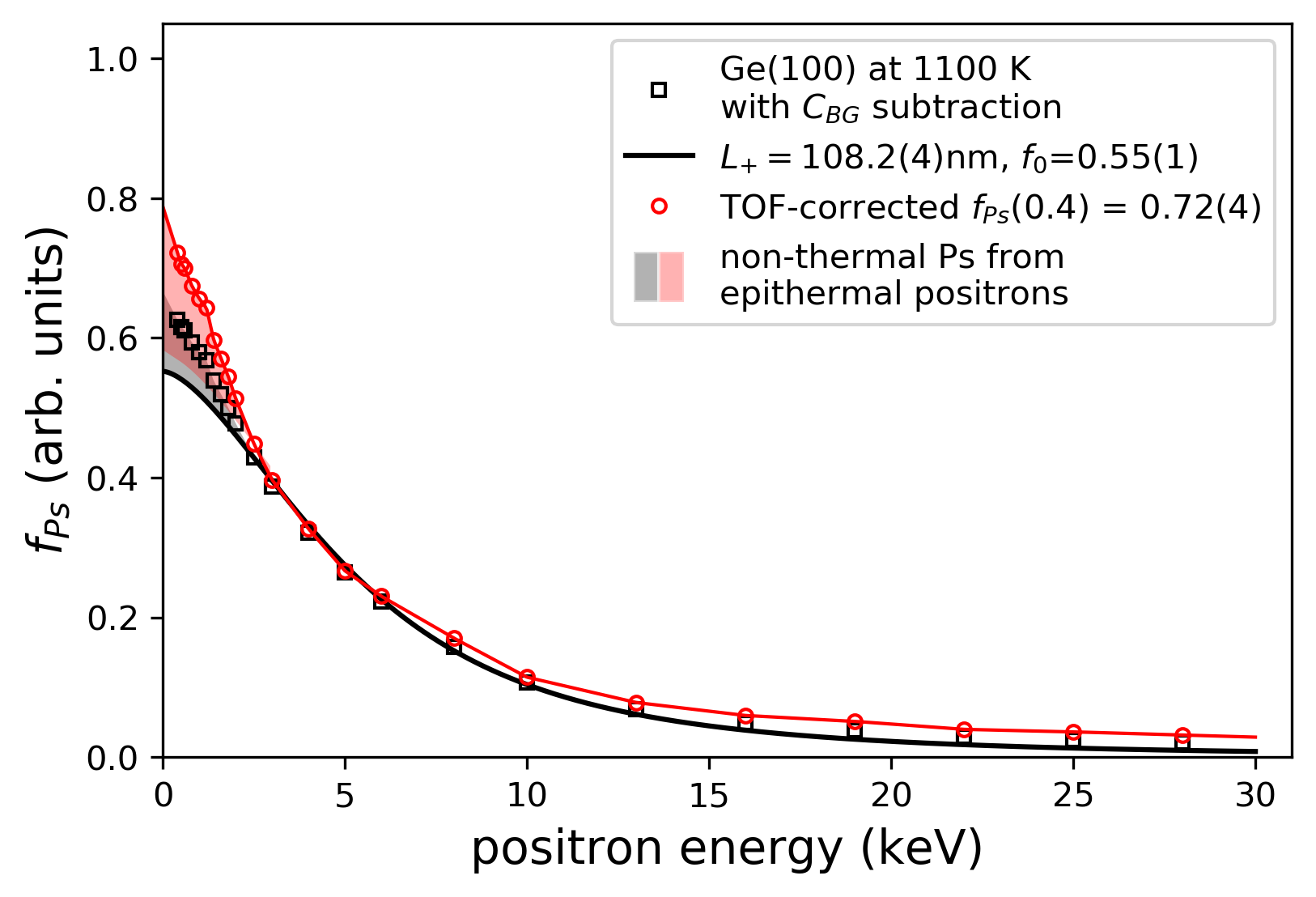}
    \caption{Ps yields resulting from superposition of simulated $2\gamma$ and $3\gamma$ spectra as demonstrated in Fig. \ref{fig:Gefitted} for all measured positron implantation energies (black squares). 
    Fitting the data by a diffusion model of thermal positrons  reproduces the expected positron diffusion lengths $L_{+}$ (solid black line). Applying a correction due to the limited flight path for o-Ps and consecutive pick-off annihilation, the absolute fractions of emitted Ps are found (red circles and a thin line as guide for the eye).}
    \label{fig:NewFGe1100K}
\end{figure}

The new $f_{Ps}(E)$ values for the energy dependent Ps formation in Ge(100) at high temperatures are given in Fig. \ref{fig:NewFGe1100K} as the black squares. 
Applying the same diffusion model as before with $f_0$ and $L_{+}$ as free parameters and only considering implantation energies greater \SI{2}{\kilo\electronvolt}, where thermal positrons validate the diffusion model, one finds diffusion parameters comparable to the ones of Fig. \ref{fig:OldFGe1100K}, but $f_0$ is reduced by about $\SI{30}{\percent}$. 
We obtain \SI{63}{\percent} Ps formation at our lowest implantation energy, which can be interpreted as a lower limit. 
About \SI{12}{\percent} of the formed Ps leaves the surface as nonthermal Ps.

The final, absolute value of $f_{Ps}$ close to zero implantation energy has to be estimated taking into account the free flight time of o-Ps after emission.
We consider this contribution by performing a time-of-flight analysis in the given geometry above the target and plug in the expected o-Ps velocities for the thermal (\SI{1100}{\kelvin}) and the nonthermal (up to few \SI{}{\electronvolt} \cite{Nieminen:1980,Nagashima2018}) component.
However, this step requires to split the obtained maximum value for $f_{Ps}$ into its thermal and nonthermal components which is demonstrated in the appendix. After this procedure, the absolute amount of produced Ps is found to be $f_{Ps}(0.4) =  \SI{0.72+-0.04}{}$, clearly contradicting the commonly accepted assumption of \SI{100}{\percent} Ps formation under the presented experimental conditions. 
The given error is systematic, since here all statistical errors are usually negligibly small. The dominant systematic error source on the absolute Ps fraction has been identified to be the Ps velocity and consecutive pick-off annihilation on nearby obstacles, which we treated in a conservative way.
A major effect leading to less than \SI{100}{\percent} Ps emission from heated germanium is the vacuum conditions; in our experiment, the pressure was \SI{5e-8}{\milli\bar}. While oxides are efficiently removed at elevated temperatures \cite{OKUMURA1998125}, this may not be the case for carbon contaminants, which could be absorbed by the surface from the residual atmosphere. Surface contaminants, however, are considered to be responsible for a not optimal Ps emission into vacuum, emphasizing the importance of a verification of the absolute fraction of emitted Ps.

\section{CONCLUSIONS}

We investigated a commonly used method for the estimation of Ps production which relies on two reference measurements with \SI{0}{\percent} and \SI{100}{\percent} Ps formation for scaling any other Ps measurement performed with a given detector system. 
In particular, the \SI{100}{\percent} reference is usually just assumed to be valid despite the experimental challenges and systematic effects this measurement might suffer from.
In this work, we presented a method without relying on this usually not justified assumption. 
A GEANT4-simulation was introduced in order to generate spectra of annihilation radiation as seen by a high-purity Ge detector in our geometry. 
These spectra were obtained for a pure $2\gamma$ source as resulting from direct positron or para-positronium annihilation, and for a $3\gamma$ source as would be produced by ortho-positronium annihilations only. The simulated detector responses were scaled up and superimposed in order to match measured spectra resulting from positrons implanted into a Ge(100) target heated to \SI{1100}{\kelvin} with varying implantation energies. 
In contrast to the common method, i.e. assuming \SI{100}{\percent} as maximum, our simulation method delivers an absolute value of $72\SI{+-4}{\percent}$ formed Ps. 
Despite this considerable difference between both approaches, at the same time the diffusion parameters for undoped germanium are reestablished.

\appendix{}
\renewcommand{\theequation}{A-\arabic{equation}}
\setcounter{equation}{0}  
\section{APPENDIX} \label{app:A}  

The measured amount of Ps consists of a thermal and a non-thermal component:
\begin{equation} \label{eq:app1}
    f_{Ps} = f_{th} + f_{nth}
\end{equation}

The thermal fraction of Ps emitted from a target of temperature $T$ is described by a Maxwell-Boltzmann velocity distribution:
\begin{equation}
    p(v) = 4\pi\bigg(\dfrac{m_{Ps}}{2\pi k_BT}\bigg)^{3/2} \, v^2 \, exp\bigg(\dfrac{-m_{Ps}v^2}{2k_BT}\bigg) \,\,\, ,
\end{equation}
where $m_{Ps}$ is the rest mass of positronium, i.e. two times the rest mass of the electron.

This velocity distribution is then converted into a time distribution, using the Jacobian determinant $\tfrac{d}{t^2}$, where $d=64\SI{+-1}{\milli\meter}$ is the mean distance to the nearest obstacles in the upper half-sphere above the target in our setup:
\begin{equation}
    p(t) = 4\pi\bigg(\dfrac{m_{Ps}}{2\pi k_BT}\bigg)^{3/2} \, (d/t)^2 \, exp\bigg(\dfrac{-m_{Ps}(d/t)^2}{2k_BT}\bigg)\cdot \dfrac{d}{t^2}
\end{equation}

The time distribution is subject to the o-Ps ground-state lifetime ($\tau_{o-Ps}=\SI{142}{\nano\second}$) which allows the calculation of the fraction $R_{th}$ of o-Ps that will annihilate into three gammas before hitting an obstacle. Since the pick-off probability for such slow o-Ps at metal surfaces is not well known, we assumed it to be \SI{100}{\percent} while using at the same time a much bigger positive-only systematic uncertainty of $+\SI{6}{\milli\meter}$ on the flight path, compensating for possible reflected o-Ps atoms. Hence, integrating overall the decaying time-distribution leads to:
\begin{equation}
    R_{th} = 1 - \int_{0}^{\infty} \, p(t) \,\, exp\bigg(-\dfrac{t}{\tau_{o-Ps}} \bigg)\, dt = 0.943\substack{+0.013 \\ -0.002}
\end{equation}

Consequently, our detector can only detect circa \SI{95}{\percent} of all thermal o-Ps atoms annihilating into three $\gamma$-quanta before it annihilates at an obstacle. 
We estimated that the mean acceptance of our detector for isotropically emitted gamma radiation along the expected flight paths of o-Ps is \SI{1.07+-0.01}{\percent}. The obtained variance of the acceptance due to different flight paths is considered negligible compared to the other uncertainty contributions.

For non-thermal o-Ps, the calculation is similar but simplified due to the assumption of mono-energetic Ps, i.e. constant velocity. 
The calculated Ps work function of Ge(100) lies around $\gtrapprox\,\SI{0}{\electronvolt}$ \cite{Bouarissa96,ZHANG:2018}. This allows to conclude that $f_{th} = f_0$.
Non-thermal Ps is therefore originating from epithermal positrons only \cite{Nieminen:1980}. Assuming a rather wide range for non-thermal Ps energies around \SI{3}{\electronvolt}, only a fraction of $R_{nth} = \SI{0.46+-0.16}{}$ can be detected by our detector before pick-off annihilation occurs at an obstacle. Due to the large uncertainty on the kinetic energy, the variance due to the unknown pick-off probability is negligible.  
The measured Ps amount $f_{Ps}$ and the fitted value $f_0$ can then be translated to an absolute value $f_{Ps,abs}$:
\begin{equation} \label{eq:app5}
    f_{Ps,abs} = f_0\bigg(\dfrac{0.75}{R_{th}} + 0.25\bigg) + (f_{Ps}-f_0)\bigg(\dfrac{0.75}{R_{nth}} + 0.25\bigg)
\end{equation}
The outcome is summarized in Table \ref{tab:summary} for a measurement at \SI{1100}{\kelvin}.
 

\begin{table}[thpb] 
	\centering
    \begin{tabular}{@{}lcccccc@{}} 
         Ge(100) &  $f_0$ &  $R_{th}$ & $(f_{Ps}-f_0)$ & $R_{nth}$ & $f_{Ps}$ & $f_{Ps,abs}$ \\
         \midrule[0.5pt] 
         \addlinespace[0.3em]
         \SI{1100}{\kelvin} &  \SI{0.55+-0.01}{} & $0.943\substack{+0.013 \\ -0.002}$ & \SI{0.076+-0.010}{} & \SI{0.46+-0.16}{} & \SI{0.626+-0.003}{} & \SI{0.72+-0.04}{}\\ 
    \end{tabular}
    \caption{Summary of Ps-components. Fractions $f_{0}$ and $f_{Ps}$ obtained from measurement can be divided by the correction factors $R_{th}$ and $R_{nth}$, respectively, in order to get the absolute fraction of emitted Ps.}
    \label{tab:summary}
\end{table}

\bibliographystyle{apsrev}
\bibliography{bib}

\end{document}